\documentclass[final,5p,authoryear,times,twocolumn]{elsarticle}

\usepackage{lineno,hyperref}
\usepackage{amsmath}
\input{Definitions}
\usepackage{nicefrac}       
\usepackage{algorithm}
\usepackage[noend]{algorithmic}
\usepackage[numbers, square, comma]{natbib}
\setcitestyle{numbers, square, comma}
\journal{NeuroImage}

\begin{document}

\begin{frontmatter}

\title{EVAC+: Multi-scale V-net with Deep Feature CRF Layers for Brain Extraction}

\author{Jong Sung Park$^1$}
\author{Shreyas Fadnavis$^2$}
\author{Eleftherios Garyfallidis$^1$}


\address{$^1$Intelligent Systems Engineering, Indiana University Bloomington}
\address{$^2$Massachusetts General Hospital, Harvard Medical School}

\begin{abstract}
 Brain extraction is one of the first steps of pre-processing 3D brain MRI data and a prerequisite for any forthcoming brain imaging analyses. However, it is not a simple segmentation problem due to the complex structure of the brain and human head. Although multiple solutions have been proposed in the literature, we are still far from having truly robust methods. While previous methods have used machine learning with structural/geometric priors, with the development of Deep Learning (DL), there has been an increase in proposed Neural Network architectures. Most models focus on improving the training data and loss functions with little change in the architecture. However, the amount of accessible training data with expert-labelled ground truth vary between groups. Moreover, the labels are created not from scratch but from outputs of non-DL methods. Thus, most DL method's performance depend on the amount and quality of data one has. In this paper, we propose a novel architecture we call EVAC+ to work around this issue. We show that EVAC+ has 3 major advantages compared to other networks: (1) Multi-scale input with limited random augmentation for efficient learning, (2) a unique way of using Conditional Random Fields Recurrent Layer and (3) a loss function specifically created to enhance this architecture. We compare our model to state-of-the-art non-DL and DL methods. Results show that even with little change in the traditional architecture and limited training resources, EVAC+ achieves a high and stable Dice Coefficient and Jaccard Index along with a desirable lower surface distance. Ultimately, our model provides a robust way of accurately reducing segmentation errors in complex multi-tissue interfacing areas of brain.
\end{abstract}


\end{frontmatter}

\begin{figure*}[h]
\centering
\includegraphics[width=1\textwidth]{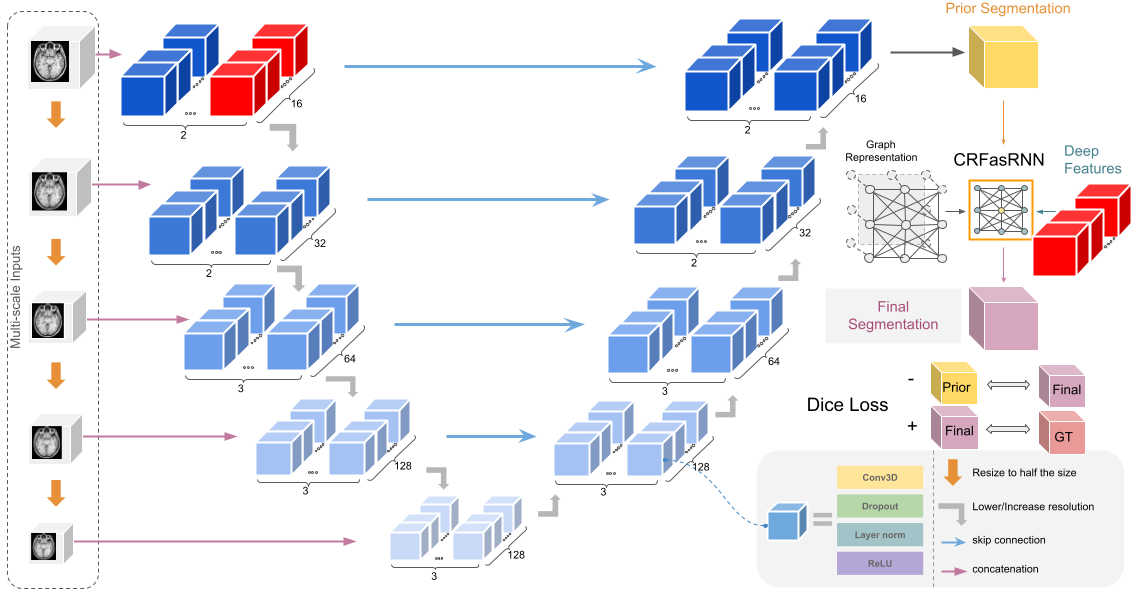}
\caption{Summary of the model architecture. The architecture uses V-net as its base model with the following important changes: multi-resolutional raw inputs (left), modified CRFasRNN (right) and additional Dice Loss (bottom right). The CRF layer uses the rear layer of the first level of the encoder (red). A negative Dice Loss is calculated between the base model's output (yellow) and CRF layer's output (purple).}
\label{fig:model}
\end{figure*}
\section{Introduction} 
  Brain MRI computes the tissue composition to create an image of the brain. However, other parts of the body also exist within the image. Unnecessary non-brain tissues could include the face, eyes, spine, etc. For additional methods such as registration, tractography and tissue segmentation to work properly, it is important to have the image of the sole brain without other parts of the body that will disrupt the algorithm or analysis. Accurate manual segmentation (gold standard) can be ideal, but it is very time-consuming with in need of an expert, who can also include biases in the results. Thus, automatic brain extraction, also called skull stripping is a semantic segmentation task necessary to pre-process the brain to perform forthcoming analyses. Despite the simple explanation of the problem, difficulties rise from several factors of the MR images. Non-brain tissues can be spatially close to the brain and have very similar intensities, especially white matter in T1-weighted images. These issues are troublesome for current algorithms. Artifacts and noise common in T1 weighted images can also disrupt the robustness of the methods.
  
  Traditional methods for skull stripping have used information about intensity ranges, shape, structure and templates in order to guide the algorithms ~\cite{bet,freesurfer,beast,alfa}. While they have proven to give a reasonable segmentation mask of the brain, they often result in large error regions that have to be resolved by manually correcting them, which can take a very long time, delaying the entire processing pipeline. Thus, there has been a large shift of the proposed methods to Deep Learning based models~\cite{isensee2019automated, hoopes2022synthstrip,salehi2017auto,lucena2019convolutional, dey2018compnet, kleesiek2016deep}.
  
  Since the use of Alexnet~\cite{krizhevsky2012imagenet} for image classification tasks, there have been many developments in using Deep Learning models for images~\cite{goodfellow2014generative,dosovitskiy2020image, he2017mask}. However, there are several reasons why it is often not optimal to implement the model architecture and ideas directly to medical images. First, is the lack of data. While it is relatively easy to acquire computer vision data that can come from various input devices, cameras, satellite images, videos etc, that is not the case for medical images. Since these contain personal information, it is difficult to gather the data in the first place and even harder to make the data public due to patient protection regulations and governmental guidelines. Hence, a lot of them cannot be used or shared among other research teams to train and evaluate their models. Additionally, labelling the data demands careful analysis of the images by experts, thus there are fewer public data with true ground truth labels. Most ground truth labels provided in the public datasets are mainly created by another machine learning method with manual editing~\cite{lamontagne2019oasis,puccio2016preprocessed,souza2018open, glasser2013minimal}. The dimension of the data is another obstacle since most of the data are 3D, and even a single T1-weighted modality image is a few hundred computer vision images in size. Thus, models suffer from heavier memory complexity problems, limiting the scale of the model. Moreover, medical imaging data tend to lack some properties Computer Vision images have (color channels, clear edge features, etc). This can limit the usage of some image augmentation techniques and complicate transfer learning, though these are both efficient tools when working with a limited amount of training data.
  
  Nevertheless, researchers have created different methods to improve DL models and allow them to work with medical images. Especially in segmentation tasks, while each model has its unique characteristics, U-nets~\cite{ronneberger2015u} and Cascade networks~\cite{schlemper2017deep} have been the skeleton of most architecture types in the field. U-net uses residual connections across multiple layers of features using different scales to take into account features of various sizes. Cascade networks exclude unnecessary information from the image by first creating a mask using a coarse network, and later using a finer network to get the exact results. Other methods have modified and enhanced these models, including but not limited to V-net~\cite{milletari2016v}, 3D-Unet~\cite{cciccek20163d} and HD-BET~\cite{isensee2019automated}. Vision Transformers~\cite{VIT} is another emerging model for medical imaging segmentation~\cite{huang2021missformer, hatamizadeh2022unetr}. While the various models have reached an overall reasonable accuracy, there is plenty of room for development in the architecture of the networks that would make these useful to clinical practice.
  
  This paper aims to improve V-net~\cite{milletari2016v}, a variant of U-net for 3D medical image segmentation. While there are noticeable changes from the original U-net architecture~\cite{ronneberger2015u}, we focus on two things which, if solved, can greatly improve its performance in brain extraction. First, the lower scale layer's information is highly dependent on features from a higher scale. Thus, at the beginning of a training phase, lower layers will not have adequate information to process. This inefficiency of training is especially problematic in brain extraction tasks since training speed is slow due to the limited mini-batch size. Second, imperfect training data labels can introduce unnecessary biases or errors for the model to learn, limiting the model's generality. Our work suggests three ways to resolve these problems: (1) multi-scale inputs, (2) a unique utilization of the Conditional Random Fields as a Recurrent Layer (CRFasRNN)~\cite{zheng2015conditional} and (3) a loss function to further remove potential errors introduced by non-DL methods that were used in creating training labels.

\begin{figure*}[h]
\centering
\includegraphics[width=1\textwidth]{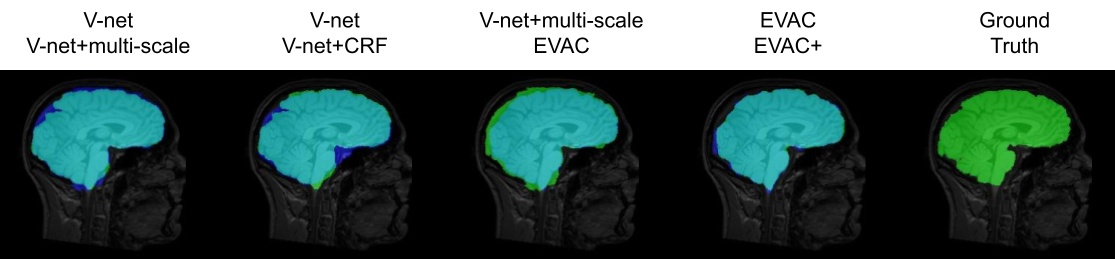}
\caption{Ablation study. Blue represents a model without a certain change, green the model with one of the proposed changes and teal the overlapping regions. Ground truth is added for reference. We can see that while both multi-scale inputs and conditional random fields do a great job of recovering False Negatives and removing False Positives, a combination of them gives a better segmentation. The additional loss function also fine-tunes the output closer to the ground truth.}
\label{fig:ablation}
\end{figure*}

\begin{figure}[h]
\centering
\includegraphics[width=.48\textwidth]{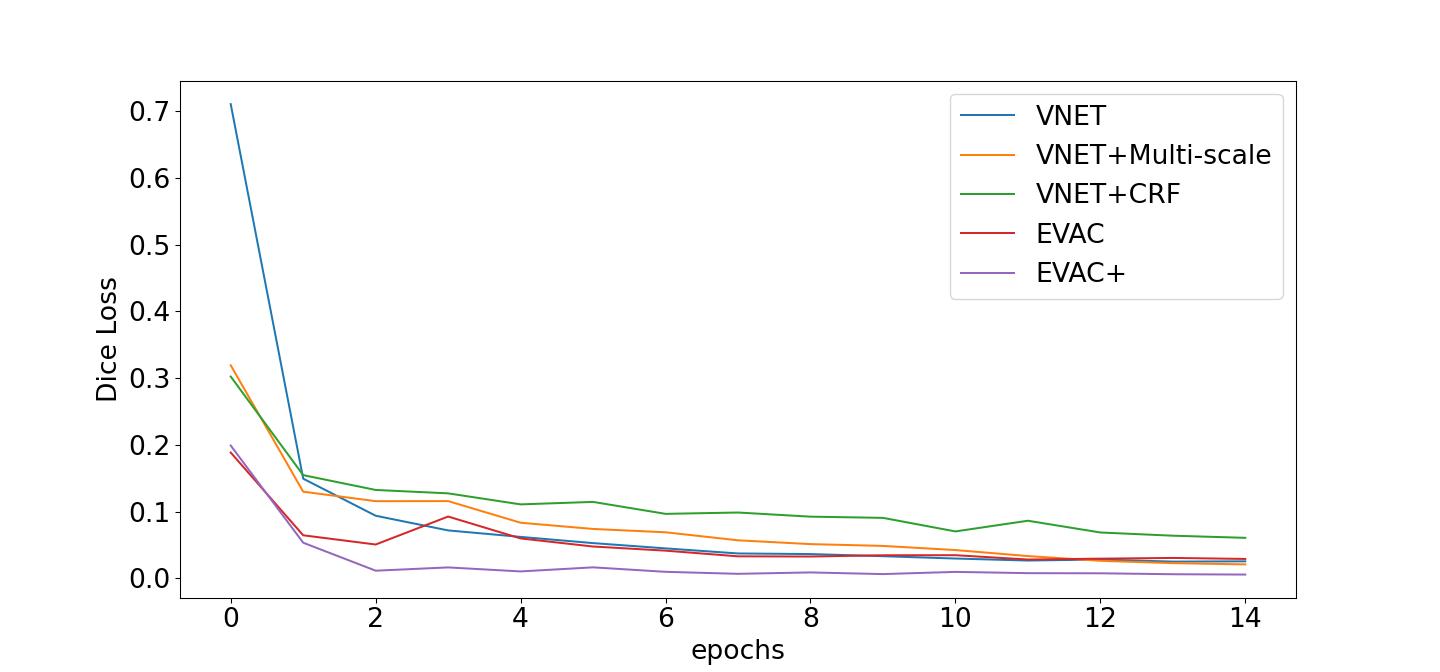}
\caption{The original Dice Loss plot. Note that the values do not include the proposed regularizing Dice Loss of our model. It clearly shows an efficiency increase in training for each improvement. Note that our model (EVAC) with the additional Dice Loss trains better even in terms of the original Dice Loss.}
\label{fig:loss_plot}
\end{figure}

\begin{figure*}[h]
\centering
\includegraphics[width=1\textwidth]{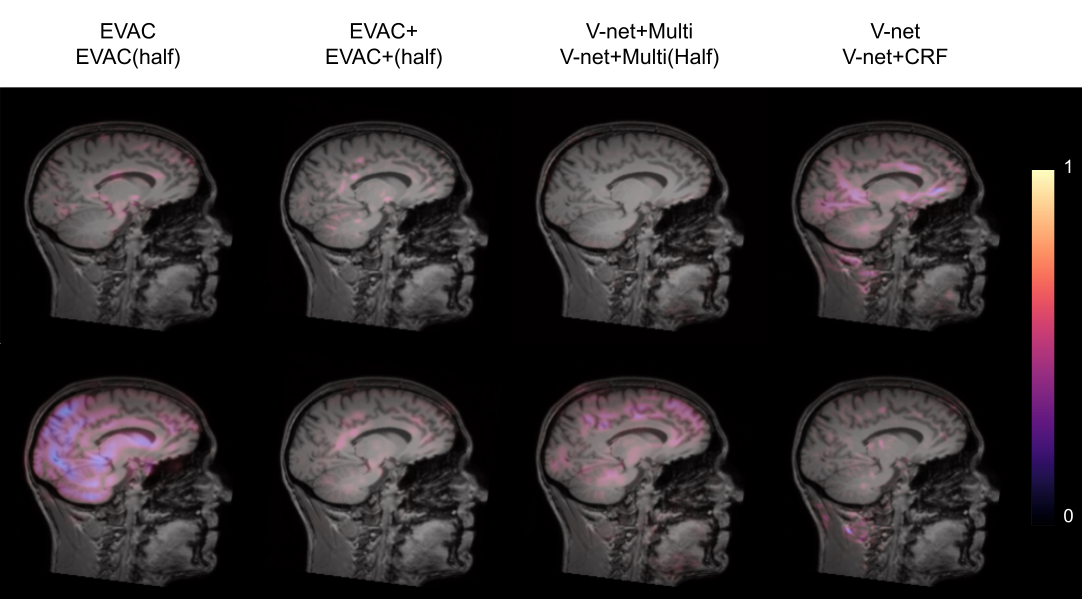}
\caption{Integrated gradients. Larger value indicates a larger gradient within that region, suggesting higher importance in prediction. Note that our model is not affected by most non-brain regions such as dura mater and neck regions. The figure also shows that EVAC+ has less important features within the brain, suggesting that the proposed CRF layer is contributing more to the model.}
\label{fig:ig}
\end{figure*}

\section{Related work}

\subsection{Brain extraction}
Despite the fact that multiple modalities can be used for brain extraction, in this paper we focus on T1 weighted brain MRI, as these are the most commonly used images for brain extraction and they are nearly always acquired with each scan.

Traditional approaches in skull stripping use known priors of the brain. They can be categorized by what type of prior the algorithm relies on: morphology, intensity, surface structure, template matching or a combination of these~\cite{review}. While every method in any category has its ups and downs, we will discuss a few methods that are popularly used by experts and work on T1-weighted images. \citet{review} provides a good summary of the various methods proposed in the literature.


Brain Extraction Tool (BET)~\cite{bet}, one of the most used methods, uses the intensity and brain surface structure as its prior. It creates an initial sphere using the intensity differences per tissue type. The sphere is deformed in each direction to form the brain's shape. Even though the model is fast and does not need any preprocessing beforehand, it has been known to have its limitations in the segmentation results~\cite{atkins2002difficulties}. 3dSkullStrip used in Analysis of Functional NeuroImages (AFNI)~\cite{afni} improves BET by adding a few modifications that help exclude the eyes and the ventricles in the image, removing some of the false positive tissues. Freesurfer's method~\cite{freesurfer} adds an intensity/structure prior, using the fact that white matter has a higher intensity in the T1 modality and is one connected segment, to create an initial volume and later deforming it to match the brain surface. ROBEX (robust, learning-based brain extraction system)~\cite{robex} uses a more hybrid approach of creating an initial mask by using a random forest classifier on extracted features, which is fit/deformed to match a more accurate surface.

Non-deformable suface-based models are widely used as well. Brain Surface Extractor (BSE) uses anisotropic diffusion filters to smooth out low-contrast edges followed by edge detection with morphological erosion. The edges are later expanded to match the brain. BEaST~\cite{beast} uses patchwise comparison to a set of templates pre-constructed from 80 segmented and manually corrected brain masks as a prior. It also uses multi-resolutions of patches to optimize the computation and reduce false negatives. ALFA (accurate learning with few atlases)~\cite{alfa} registers manually segmented neonatal brain atlases to the target image, selects the few closest to the average of them and fuses the labels using machine learning approaches to create the segmentation output.


Various Deep Learning approaches have been considered in brain extraction over the years~\cite{isensee2019automated, hoopes2022synthstrip,salehi2017auto,lucena2019convolutional, dey2018compnet, kleesiek2016deep}. One of the major factors that characterize these models is the dimension the data is treated as. For Deep Learning models in brain extraction tasks, it might be troublesome when trying to train a model that stores all the full 3D features in memory. Most 2D convolutional network methods use 2D slices from all planes to get segmentations from each plane, later merging the results~\cite{henschel2020fastsurfer,lucena2019convolutional,salehi2017auto}. Dismantling the 3D image into smaller cubes has also been considered~\cite{kleesiek2016deep}. Other approaches have cropped the image to a smaller size to use it as a form of augmentation as well~\cite{isensee2019automated, nobrainer}. In our work, we chose to use the 3D image as a whole, decreasing the resolution of the input (EVAC uses a 2mm cube per voxel) to a usable dimension.

Models that work with damaged brains and multiple modalities have also been proposed. BrainMaGe~\cite{thakur2020brain} was trained to be more specific to brains with tumors. HD-BET~\cite{isensee2019automated} was trained with brains with glioblastoma along with healthy subjects from public datasets to provide a generalized brain extraction method. Both were created to work with any of the 4 structural modalities. Synthstrip~\cite{hoopes2022synthstrip} created synthesized images by augmenting not just the image itself but on each part of the known whole-head anatomical segmentation. The synthesized images were used to train the model to be generalizable to multiple modalities. 

While there are multiple types of architectures proposed in the literature, this paper will focus on one of the most utilized types of models, U-net. The popularity of U-net comes from its clever architecture of forcing the model to learn multi-scale features by using a max pooling layer for every block of layers and using an autoencoder-like structure with skip connections. These skip connections between the encoder and decoder part of the model add an additional pass of information on each scale, making the model more segmentation friendly than classic autoencoders.

Since U-net was introduced, many DL methods that train on medical images now use it as their base architecture~\cite{salehi2017auto,lucena2019convolutional,isensee2019automated,hoopes2022synthstrip}. Though the base model was created for 2D medical image segmentation, there have been a few models that modify this architecture to work for 3D \cite{cciccek20163d,milletari2016v}. Other improvements have been introduced as well. \citet{li2019cascade} added cascade networks to U-net to help segmentation on data where the background might be significantly major than the foreground. W-Net~\cite{xia2017w} used two U-net structures to do unsupervised segmentation of images. HD-BET~\cite{isensee2019automated} weighted the complexity of the model to be heavier in the encoder along with loss calculation for each level of the decoder to facilitate training.

\subsection{Conditional Random Fields}

CRFs were originally used for segmenting and labelling sequence data~\cite{lafferty2001conditional}. The method is based on assuming the label's probability distribution to be a Markov Random Field (MRF), e.g. affected by only its neighbors. Various approaches have been conducted to use CRFs in image segmentation. The major difference in these methods are: a) the initial segmentation algorithm for the prior distribution (probability of a pixel belonging to a certain label) and b) how the graph is constructed. \citet{shotton2009textonboost} used texture/color/spatial information to create an initial map with 4 neighboring pixels for the graph structure. \citet{fulkerson2009class} created superpixels, a small group of pixels, and calculated the histogram of them to define the initial prior. Enhancing the CRF itself was also suggested using a hierarchical strategy~\cite{ladicky2009associative,kumar2005hierarchical}.

\citet{krahenbuhl2011efficient} suggested an efficient way of calculating CRFs so that the model could use a fully connected graph instead of just neighboring pairs. Their method uses the position and intensity information of the pixels to create an energy function, which is minimized to reduce the difference between the same labels. The initial segmentation is done by following the feature extraction concept from \citet{shotton2009textonboost}. We will explain more about the theorem behind their model.

Let us consider $X = {X_1, X_2, ..., X_N}$ to be a vector of label assignments on each pixel $1$ to $N$, where the value of each element is within the possibilities of the label. In our case, since brain extraction is a binary segmentation problem, it would be either 0 or 1. Since we are assuming an MRF, it can be approximated as a Gibbs distribution with neighboring interactions. Thus, given image $I$, we can write a Gibbs distribution $P(X=x|I) = \frac{exp(-E(x|I))}{Z(I)}$. $Z(I)$ is the denominator dependent on only the image itself, so can be ignored. The model aims to find the maximum $x$ that has the highest probability, thus the lowest energy function. \citet{krahenbuhl2011efficient} defines the Gibbs energy as
$$E(x) = \sum_{i}\psi_{u}(x_i) + \sum_{i<j}\psi_{p}(x_i, x_j)$$
where $\psi_{u}$ is the unary potential (potential of a single pixel/voxel) and $\psi_{p}$ is the pairwise potential (potential from its neighboring information). Note that the dependency on the image $I$ was omitted for convenience. In our case, the unary potential will be our base architecture's output. The pairwise potential is defined as
$$\psi_{p}(x_i, x_j) = \mu(x_i, x_j)\sum_{m=1}^Kw^{(m)}k^{(m)}(f_i, f_j)$$
where $\mu$ is the compatibility function that serves as a penalty between different labels with similar features and $k^{(m)}$ is the kernel that utilizes feature $f_i, f_j$ information. The paper uses position (smoothness) and intensity/position (appearance) kernels as the feature information. They are defined below.
$$k(f_i, f_j)=w^{(1)}exp(-\frac{|p_i-p_j|^2}{2\theta^2_{\alpha}}-\frac{|I_i-I_j|^2}{2\theta^2_{\beta}})+w^{(2)}exp(-\frac{|p_i-p_j|^2}{2\theta^2_{\gamma}})$$
$p_i$ and $I_i$ are each the spatial and intensity information of pixel i. $\theta_{\alpha},\theta_{\beta},\theta_{\gamma}$ controls the degree of each feature.

They also provide a way of efficiently approximating the posterior distribution (essentially the probability map) using mean field approximation. A quick summary of the method is described here. It assumes an independent relationship between the latent variables, in our case the true labels of each pixel/voxel. Since our goal is to find the most accurate approximation of the posterior, we can aim to find the minimum KL-Divergence between the posterior and the approximation. Through Bayesian inference, one can calculate that the prior is a sum of the KL-Divergence and negative variational free energy. \citet{krahenbuhl2011efficient} showed that this can be written in the form of the unary potential and the pairwise potential. The assumption on the independence of the latent variables lets each variable be updated separately treating the other variables as constants. Thus, each iteration of the latent variable update that increases the negative variational free energy is guaranteed to decrease the KL-Divergence. Optimally, the converged output would be the true segmentation probability map of the image.
Details on the method can be found in \citet{krahenbuhl2011efficient}.


The Deep Learning model's image segmentation output is mainly a softmax result of the final layer. Since this gives a probability distribution of a pixel belonging to a certain label, using CRFs after a Deep Learning model would be a good way of improving the final result beyond the base model. CRFasRNN~\cite{zheng2015conditional} integrated the iteration process from \citet{krahenbuhl2011efficient} as a Recurrent Layer. Unlike the original, fully connected CRF model including the CRF process in the model architecture itself increases optimization efficiency and reduces the number of hyperparameters. Since then there have been suggestions to improve the model~\cite{arnab2018conditional}, but not for volumetric medical images. While both the manual and Recurrent Layer approaches have been utilized in many medical imaging domains~\cite{lesion_crf, fu2016deepvessel, wang2020unified, li2018cancer, duy2018accurate, magnano2014conditional}, it has been suggested that using naive CRFasRNN layer within the Deep Learning architecture does not help much with the volumetric segmentation~\cite{monteiro2018conditional}.

The reasoning behind this lack of performance could be several reasons, but in this work, we focus on the single-channel problem. When CRFs are used in Computer Vision, the feature space is constructed using color channels. However, many medical images do not have multiple channels. Thus, the CRF algorithm will not have enough information to refine the initial segmentation output.

\begin{figure*}[h]
\centering
\includegraphics[width=1\textwidth]{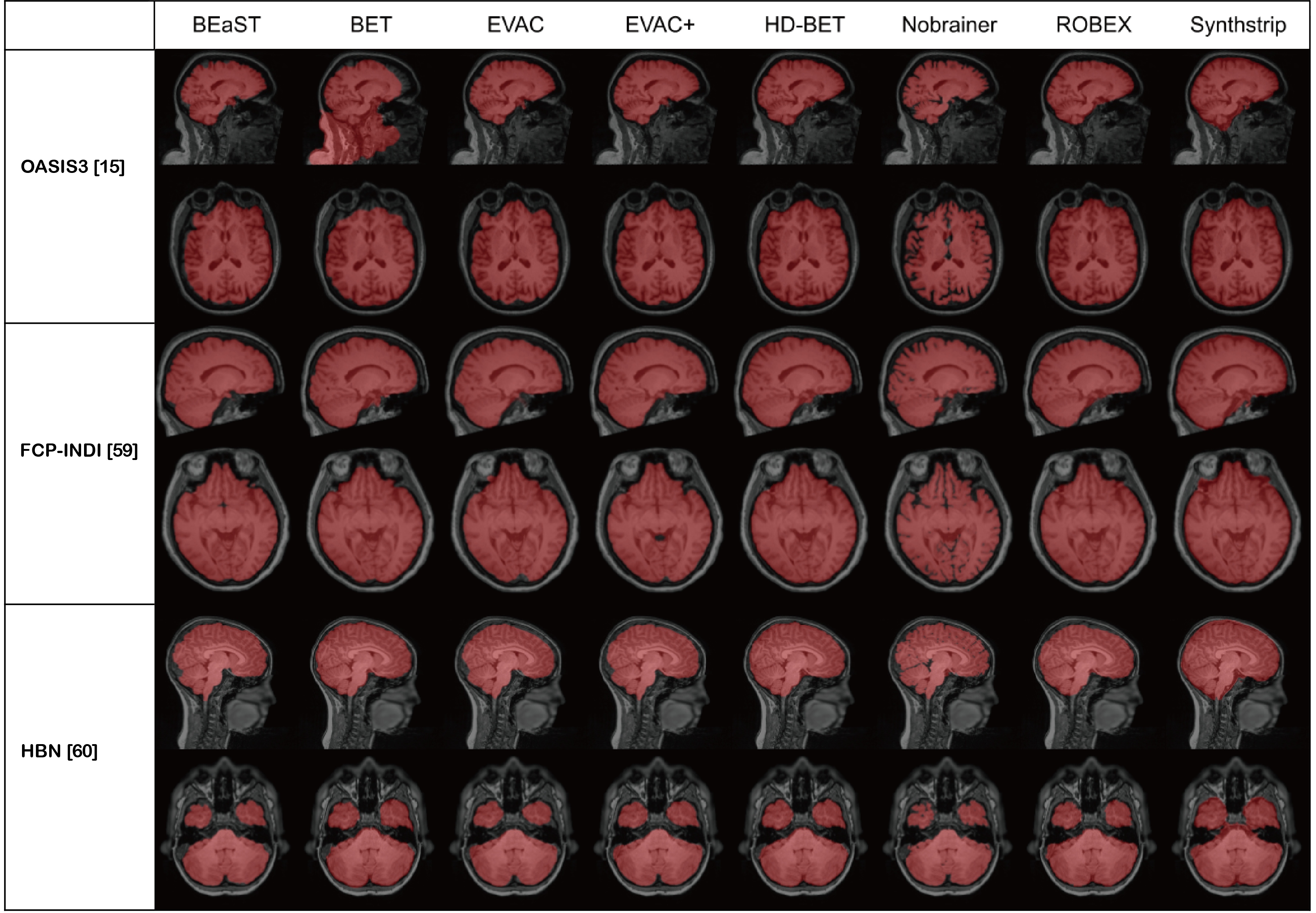}
\caption{Qualitative robustness analysis of the compared methods. The images are from OASIS, Parkinson's dataset from the FCP-INDI project and pediatric brains from the Healthy Brain Network dataset. Red marks the segmentation results of each method.}
\label{fig:robustness}
\end{figure*}

\begin{figure*}[h]
\centering
\includegraphics[width=1\textwidth]{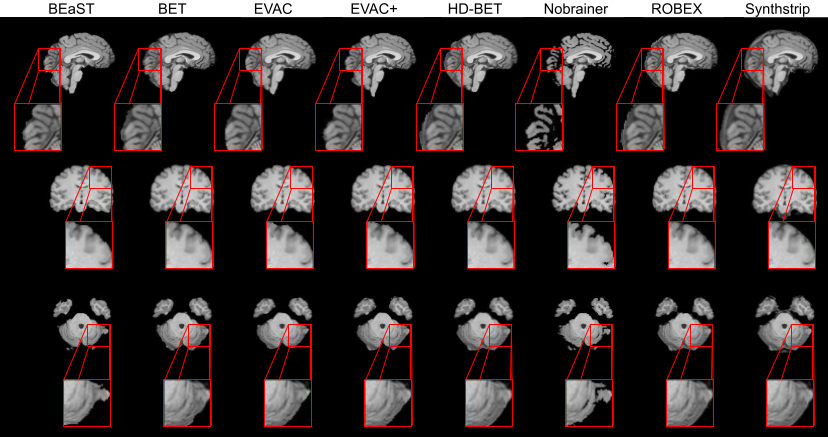}
\caption{Comparison between our models with (EVAC+),  without the additional Dice Loss (EVAC) and other established state-of-the-art models. Specific regions were zoomed in to emphasize the improvements our model is achieving. Results show that our models get an accurate local segmentation on the surface, whereas most of the other methods either under-segment or include the dura mater. Similar accuracy improvements are also visible near the central sulcus and the cerebellum. T1 images from the IXI dataset.}
\label{fig:comparison}
\end{figure*}

\begin{figure*}[h]
\centering
\includegraphics[width=0.8\textwidth]{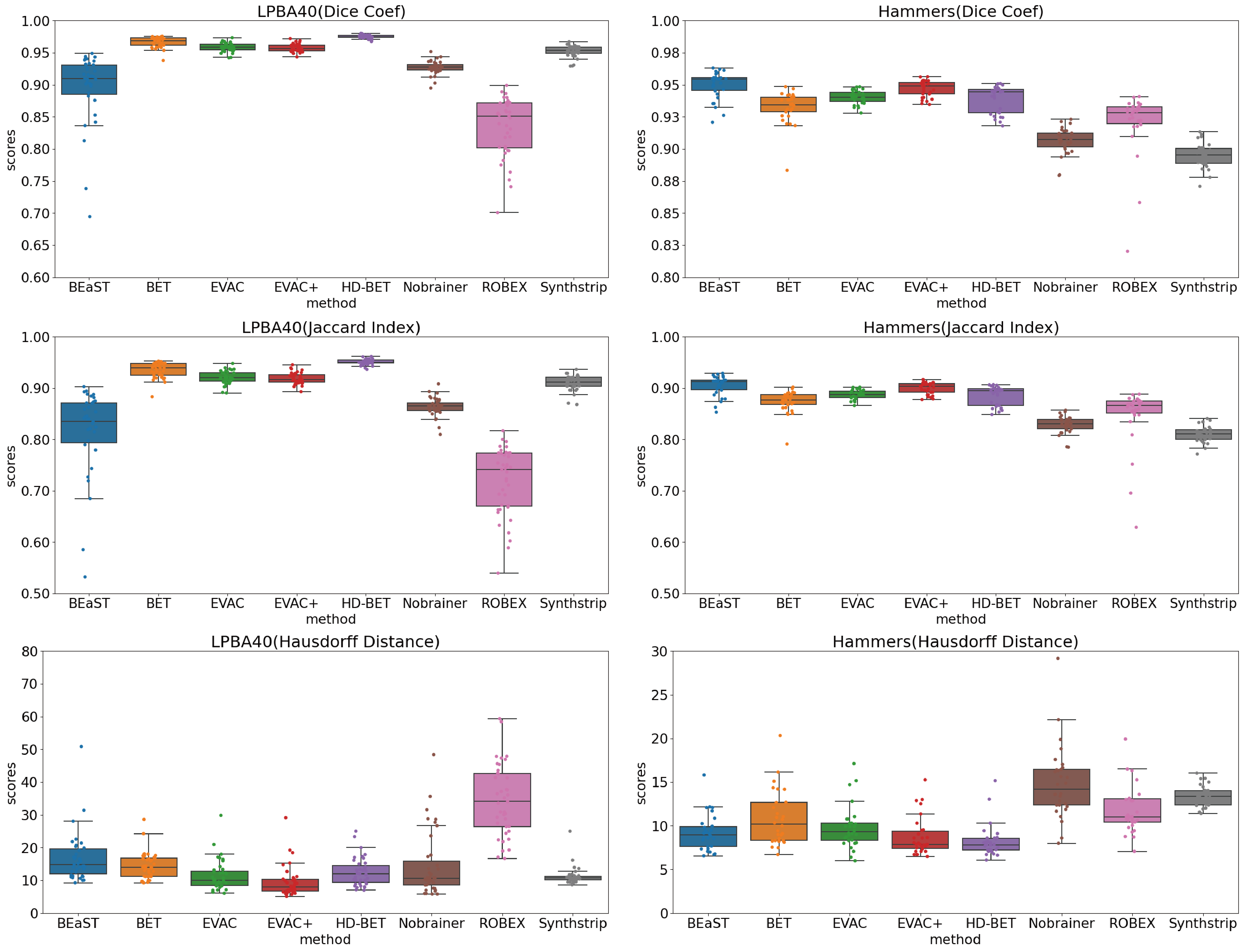}
\caption{Evaluation of our models compared to other methods. The metrics are Dice Coefficient, Jaccard Index and Hausdorff Distance (top to bottom). LPBA40 dataset and Hammers Atlas dataset (left to right) were used as the test sets. Our models with (EVAC+) and without the additional Dice Loss (EVAC) have a stable near-top accuracy in both datasets and metrics, while others have either lower scores in a dataset or unstable results.}
\label{fig:plots}
\end{figure*}

\begin{figure}[H]
\centering
\includegraphics[width=.47\textwidth]{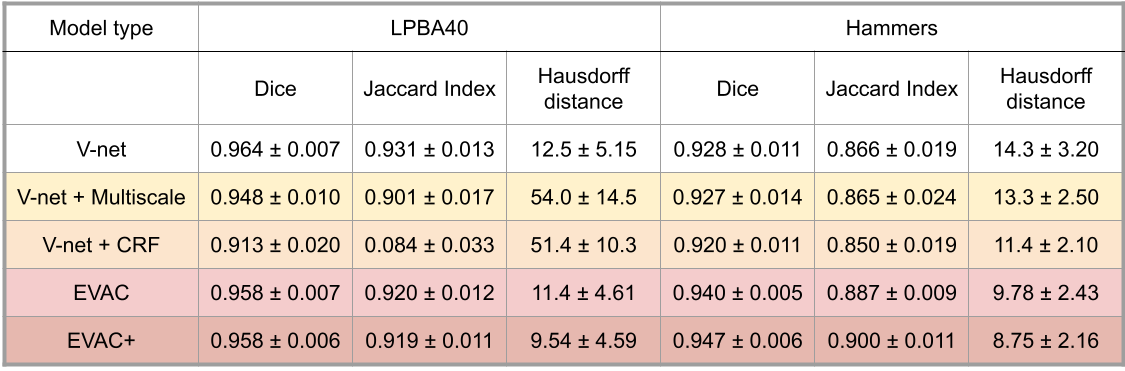}
\setcounter{figure}{0}
\renewcommand{\figurename}{Table}
\caption{Quantitative ablation study of our model. EVAC and EVAC+ on datasets LPBA40 and Hammers. Note that Hausdorff distances clearly improve with both datasets.}
\label{fig:metric_table}
\end{figure}

\section{Methods}

Due to the size and complexity of brain MRI and the lack of accurate public data, it is incredibly important to have a robust and efficient model that works with limited resources. This paper will propose three ways to enhance Deep Learning architectures for brain extraction: (1) multi-scale inputs for the efficiency of the model, (2) a distinctive usage of the Conditional Random Field Recurrent Layer and (3) a matching loss function to get a finer segmentation result. The overall architecture is shown in Fig.~\ref{fig:model}.

\subsection{Multi-scale}
We first propose feeding raw multi-scale inputs to the model, as previously suggested by other segmentation models\cite{mehta2017m, xu2018multi}. We add the lower-scale inputs by concatenating them to the output of each downsampling convolutional layer as shown in Fig.~\ref{fig:model}. We revisit the reasoning behind this as it has a synergistic effect with our other improvements. 


We can write a general equation for the output of a layer in the model as:
$$y = \sigma(z)$$
$$z = wx + b$$
where $w$ is the weight, $x$ is the input, $b$ is the bias and $\sigma$ is any non-linear function. The gradient of $w$ is calculated through back-propagation to update the weight (kernel in convolutional layers). By the chain rule, we know that the gradient value is a multiplication of the gradients of the deeper layers, the non-linear function and the input to the current layer. We can write this as 
$$\frac{\delta L}{\delta w} = G\frac{\delta x^{\prime}}{\delta z}\frac{\delta z}{\delta w}$$
where $L$ is the loss function, $G$ is the previous gradients in the chain rule, $x^{\prime}$ is the input to the next layer. There are two major changes one can make in the layer to modify the target gradient. One is altering the non-linear function, hence changing $\frac{\delta x^{\prime}}{\delta z}$. We aim to change $\frac{\delta z}{\delta w}$ which is the input to the layer. In other words, we are defining $x$ to be
$$x = x_{prev} + x_{raw}$$
$x_{prev}$ stands for the input from the previous layer that reduces the feature size. This is present in any U-net-type architecture. $x_{raw}$ is the raw input resized to match the layer's feature shape. It forces the weight to update using meaningful information even if the $x_{prev}$ provides a non-important feature, important especially when training from a randomly initialized model.

The lower scale inputs will ultimately reinforce the multi-scale scheme of V-net so that each end of the encoder structure will contain more accurate corresponding scale's features. We utilize this in our next proposed change.

\subsection{Enhanced CRFasRNN}
While the multi-scale inputs provide good results alone, they still contain some bias from the non-human-made labels. We need our output to follow the complex, intricate and detailed structure of the brain, which is often not captured in the ground truth. Many features are considered when calculating a brain mask, but one cannot deny the fact that continuity in intensity and structure is an important factor. As CRFs are known to minimize the discrepancies in labels between neighboring pixels with similar intensity and spatial information, they can be used to create a fine-grained segmentation result.


As suggested in \citet{monteiro2018conditional}, we add a Recurrent Layer that essentially does the matrix factorization optimization iteration step of the CRF. The layer is added right after the output of the multi-scale model part of the network. We have tried both training from scratch and fixing the weights of the main architecture through training and later only changing the weights of the Recurrent layer. Though fixing the weights can prevent any unnecessary change in the close-to-optimal parameters of the main model, training all the weights from the start provided better results. This is possibly due to the fact that the previous model already learned a stable local minimum and just the recurrent layer is not enough to resolve this. We used CRFasRNN Layer as provided in \citet{monteiro2018conditional} and \citet{adams2010fast}, which provides an efficient way of calculating the kernel features. Instead of 10 iterations recommended in the previous studies, we found 20 iterations to give much better results, which can be decreased back to fewer iterations during prediction.

However, this does not solve the single channel problem mentioned above. Thus, one of our major contributions is adding learned features within the network along with the image intensity to improve the CRF layer. Specifically, we used the last layer of the highest scale output of the encoder in the model. While our ablation study below will show that it alone has a beneficial effect on the model's performance, the more accurate features created from the multi-scale inputs have a synergistic effect with our modified CRFasRNN layer.

\subsection{Loss for CRFasRNN}
While in practice, we also encountered problems where sometimes the CRFasRNN layers fail to significantly change the model output, even when trained together from scratch on certain areas more stable intensity-wise than the others. Hence, we propose adding a second loss function, negative Dice Loss between the original model's output and the CRF layer's output. Given a good loss weight, this can enforce the base model to give a more stable crude segmentation while the CRFasRNNlayer improves the segmentation result by a reasonable amount. We can write the new loss function as this

$$D(y, y') - \lambda D(z, y)$$
Where $D$ is Dice Loss $y$ the final prediction, $y'$ the provided ground truth, $\lambda$ the additional loss weight and $z$ the prediction before the Recurrent layer.

\textbf{Additional processing:} For all the methods tested, including our model, we have removed small segmentation errors that might be present in the result. Using prior information about the brain structure, we have limited the segmentation result to the largest connected component, while filling the black spaces within the segmentation.

\subsection{Training details}
Our models were trained using two Tesla V100s. All the models compared in the ablation study were trained until convergence. More specifically, EVAC, EVAC+, V-net with CRF layer, V-net with multi-scale inputs and base V-net were respectively trained for 18, 18, 18, 32, 40 epochs. Learning rate of 0.01 was used. Dropout rate of 0.2 was used for the initial layer and 0.5 was used for the rest.

1438 T1 MR images were used for training from the following public datasets: the Human Connectome Project(HCP)~\cite{glasser2013minimal}, CC359~\cite{souza2018open} and NFBS~\cite{puccio2016preprocessed}. The provided labels of the datasets were used for training. For testing purposes, the LPBA40 dataset~\cite{lpba} and adult brains from the Hammers dataset~\cite{hammers1, hammers2} were used. To show the performance as images, IXI dataset~\cite{ixi} was used. Note that unlike many state-of-the-art methods, it was trained only on public datasets with minimal augmentation sampling, described below. DIPY Horizon~\cite{garyfallidis2014dipy} was used for visualization.


Each image was loaded to a common space of 1mm cube per voxel using the affine matrix accompanied with the file. The image was translated so that the center of the image would be the voxel coordinate of (128, 128, 128). Image was padded or cropped to have a size of (256, 256, 256). No additional registration to template process was required. 

Augmentation has been known to force the model to learn structural components. It is particularly important for medical images due to the lack of data. The image was first normalized to a range of 0 to 1. For intensity augmentation, we have used scaling (range of 0.9 to 1.1) and shifting (range of -0.1 to 0.1) of values to add/multiply random noise. For transformation augmentation, random rotation (maximum 15 degrees) and translation (maximum range of 10mm) were used. This was to compensate for most of the minor differences T1 weighted images might have when in the same coordinate space. On every epoch, a single random augmentation method was chosen for each image. We only trained on the augmented images, i.e. did not increase the size of the training dataset per epoch. This was done to improve generality without increasing the time complexity of training. We would like to emphasize EVAC+ was able to work with the more extreme range of augmentations while models without the modified CRFasRNN layer could not.

More details on the training environment and the exact parameters for the model architecture can be found in our code \\ \url{https://github.com/pjsjongsung/EVAC}.

\section{Results}
Evaluation between models was done with three metrics. The Dice Coefficient and Jaccard Index were selected to measure the similarity of the prediction to the ground truth. While both are used for the evaluation of image segmentation, Dice Coefficient adds more weight to the True Positives than Jaccard Index. Hausdorff Distance was used to measure the surface distance error. Higher scores are better for Dice Coefficient and Jaccard Index. For Hausdorff Distance, lesser scores are better.

In the figure descriptions, EVAC+ refers to our model with the additional loss. V-net+multi refers to a V-net model with the multi-resolution scheme and V-net+CRF refers to V-net with the CRFasRNNLayer but without the multi-scale inputs.

\subsection{Ablation study}
We first compare results from V-net~\cite{milletari2016v}, V-net with multi-scale inputs or CRFasRNN layer, EVAC, and EVAC+ with the proposed loss function to show the development of results per each change in the model. Fig.~\ref{fig:ablation} provides a summary of what each enhancement to the model does in terms of output. Ground truth image is also provided in the figure for overall performance evaluation. 

Each improvement (multi-scale input, the proposed CRF layer and loss function) of the architecture corresponds to an improvement of accuracy. The multi-scale input enforces the model to use large-scale features, which is missed in the V-net output. However, this also introduces non-brain regions to the segmentation result. The proposed CRF layer corrects most of the false positives. However, it is not completely free from the base model having major influence over the CRF layer. We can see that the new loss function projects additional importance to the CRF layer, inducing greater correction in the segmentation output. To show that the proposed CRF layer is beneficial even without the multi-scale scheme, we compare V-net like model with the same model plus the CRF layer. You can see that it oversegments some regions, but recovers important brain regions as expected.

The improvement is not only in the accuracy but the efficiency of training. Fig.~\ref{fig:loss_plot} shows how the loss changes with the number of epochs. Note that the training environment is identical between models except the proposed changes (e.g. CRF layers, additional loss). The plot clearly shows the learning efficiency increase per enhancement of the model. Another important phenomenon is that even though the regularizing Dice Loss is applied at the model level, it still leads to faster convergence of the original loss.

We further show integrated gradients~\cite{ig} to provide an explanation of how our changes are improving the model. Fig.~\ref{fig:ig} shows that our changes lead the model to focus less on insignificant regions within the image. Comparison between EVAC and V-net+multi-scale suggests that our proposed CRF layer pushes the model toward better usage of multi-scale inputs. The significant difference in feature importance within the brain between EVAC and EVAC+ is possibly due to the CRF layer's refining step playing a major role, thus needing less focus on low level features. This can be beneficial when dealing with abnormal or noisy images.

Quantitative results on the test dataset is shown in Table~\ref{fig:metric_table} to emphasize the synergistic effect. While the proposed changes individually do not help the model's accuracy, EVAC and EVAC+ in the end manage to get lower Hausdorff Distance, indicating that the proposed changes has a positive effect in refining the cortical surface regions. 

We would like to emphasize that besides some minor details (e.g. number of epochs until convergence) the models were unchanged except for the major proposed changes. Thus, the increase in accuracy would have been purely from the improvement of the model architecture.

We also show Fig.~\ref{fig:robustness} to emphasize the robustness of our methods. The images are from the OASIS3 dataset~\cite{lamontagne2019oasis}, Parkinson's from FCP-INDI~\cite{badea2017exploring} and pediatric brains from the Healthy Brain Network dataset~\cite{richie2022analysis}. It is evident that our method is robust in a wide area of clinical/pediatric data, even though such data were never introduced during training.


\subsection{Comparison}
We compare our EVAC model to other publicly available state-of-the-art methods. The models were chosen based on the accessibility. For Deep Learning models, only pre-trained models were chosen for two reasons. First, the training dataset is a critical part of DL methods in brain extraction. Thus, it is not reasonable to train the models again with a fixed dataset. Also, since we are planning to release the pre-trained model for actual use, it is more reasonable to compare to the models that are provided for the same purpose.

Fig.~\ref{fig:comparison} show the results of each model on an image from the IXI dataset. Fig.~\ref{fig:plots} shows the average of each metric on the test dataset. Both the EVAC model with and without the proposed loss function were included to emphasize its effects. While most methods fail on removing the dura mater on the surfaces of the brain or over-segment, both of our models are mostly free from that problem. BEaST is another method that is free from this issue but is known to have problems around the cerebellum, which our models do not. Additionally, by using the additional loss function, our model returns segmentation outputs with higher detail.

\section{Discussion (Limitations and Future work)}

Our model provides a robust and efficient DL model for brain extraction with limited data and augmentations. Results clearly show the stable accuracy compared to other methods, especially in the cortex/dura mater interface. The ablation study performed suggests that Deep Feature CRFasRNN layer with multi-scale inputs and negative Dice Loss have synergistic effects. The improvement can be seen not only in accuracy but also in efficiency during training.

Despite our model's performance in the evaluation dataset, it is limited to T1 weighted MRI data unlike other proposed methods~\cite{isensee2019automated, hoopes2022synthstrip, thakur2020brain}. However, we believe when provided with enough variety in images, the model would be able to train up to a state where it can provide output for abnormal brains, as the model utilizes features from the image, which is more robust to the type of image. The additional loss would also help correct local issues with the segmentation. Further work would be extending the dataset for general use by adding more variety in modalities or clinical cases of the training data.



For the past few years, BEaST had the top accuracy among both traditional and Deep Learning methods. However, the time complexity of the method has prevented it from being widely used in the field. Thus, even though it is often problematic, BET has been the top-used Skullstripping method. Our method however clearly shows comparable or higher accuracy and does not suffer from such problems as it is a Deep Learning model. Therefore, we believe our method could be not just helpful for the medical image segmentation community but all fields that could benefit from an accurate Brain Extraction.

The paper emphasizes the enhancement of the overall architecture, but keeps the changes in the base architecture minimal. However, there are various changes we can make to the base V-net architecture. Instance Normalization was proposed in StyleGAN~\cite{karras2019style} and had cases where it performed reasonably well in medical imaging  tasks~\cite{schilling2019synthesized}. Many losses have been proposed as alternatives to Dice Loss~\cite{salehi2017tversky, hoopes2022synthstrip}. Even though we are not using the whole Vision Transformer~\cite{VIT} architecture, it has been reported that just applying self-attention to the Convolutional Neural Networks could be beneficial to its robustness~\cite{wang2022can}. Future work would be refining our model with these techniques for better quality outputs.

It is still controversial on whether a segmentation model should be based on Vision Transformers or Convolutional Networks. Despite the recent advancements in Vision Transformers, we did not focus on or use the method in this paper as they tend to be trained on more training data and training epochs~\cite{bai2021transformers}, which is often not suitable for large volumetric medical images. Also, despite the efforts to combine U-net and Transformers to utilize more global contextual info~\cite{jiang2022swinbts}, our model focuses more on correcting local errors, which is where the errors from methods are in Brain extraction tasks.

Despite the fast speed of the Deep Learning prediction using GPUs, CRFasRNNLayer with more features can be a bottleneck when done with CPUs. While we can reduce the time complexity by reducing iterations within the CRF layer as the original paper suggests, training with simpler models can be a faster solution. Future work would be retraining the outputs of the model with a simpler model to keep the benefits of the CRF layer's output refinement while reducing time complexity.

\section{Conclusion}
Removing non-brain tissues from MRI data is a necessary step for nearly all forthcoming analyses of these images. Various Deep Learning approaches have been considered in the literature. Here we propose EVAC+, a new DL architecture that combines a V-Net architecture with CRF layers and multi-scale inputs. Results show that the proposed CRF layer with a negative Dice Loss for refining the segmentation results can significantly improve the outcome. The improvements are especially profound in cortical surfaces, which most methods fail to achieve a precise segmentation. We also show that this ready-to-go approach can efficiently reach higher accuracy with fewer epochs. We think it is exciting and refreshing that a revisit to CRFs in medical imaging can be used in tandem, creating a compound effect with Neural Networks, and further improve crucial medical imaging problems. In addition, we devised an efficient approach for training EVAC+ using limited resources. The code will be open-sourced via DIPY~\cite{garyfallidis2014dipy} to disseminate its use to the wider community.

\section{Data and Code availability}
Code and tutorials for EVAC+ are publicly available in GitHub and can be found here \\ 
\url{https://github.com/pjsjongsung/EVAC}.

Code is written using Python language with dependencies in DIPY and Tensorflow. The CRFasRNN layer skeleton was adapted from \\\url{https://github.com/MiguelMonteiro/CRFasRNNLayer}.

EVAC+ will also be made publicly available through DIPY~\cite{garyfallidis2014dipy} at \url{https://github.com/dipy/dipy}.

All data used in the article are publicly available.

\section{Acknowledgements}

We would like to acknowledge that research reported in this publication was supported by the National Institute Of Biomedical Imaging And Bioengineering of the National Institutes of Health under Award Numbers R01EB027585 and 2R01EB017230-05A1.

We are also grateful to Indiana University for providing access to computing clusters Bigred200 and Carbonate systems with NVIDIA/TESLA GPUs for training and testing our method.

Finally, we would also like to thank the Graduate Program in Intelligent Systems Engineering and Program in Neuroscience of Indiana University, Bloomington for sponsoring Jong Sung Park.

\section{Competing Financial Interests}
The authors declare no competing financial interests.




\biboptions{authoryear}
\bibliography{mybibfile}

\end{document}